\documentclass{moriond}

\def\be{\begin{equation}}
\def\ee{\end{equation}}
\def\bea{\begin{eqnarray}}
\def\eea{\end{eqnarray}}

\newcommand{\Photo}{\includegraphics[height=35mm]{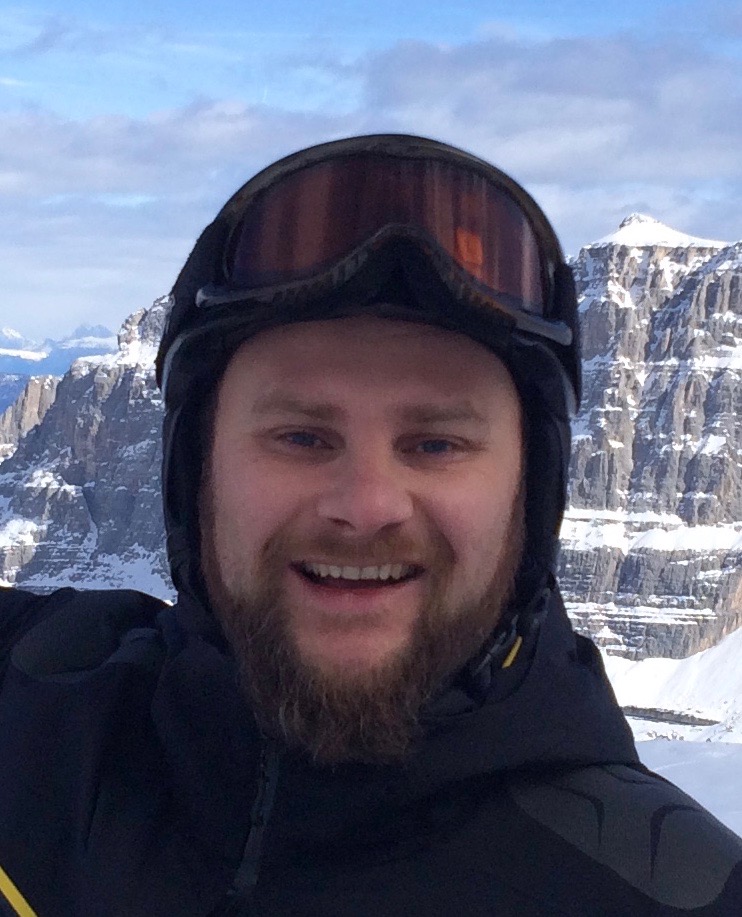}}

\begin{document}
\vspace*{4cm}
\title{GLOBAL DYNAMICS FOR NEWTON AND PLANCK}

\author{ALEXANDER VIKMAN}

\address{CEICO--Central European Institute for Cosmology and Fundamental Physics,
\\
FZU--Institute of Physics of the Czech Academy of Sciences, \\
Na Slovance 1999/2, 18221 Prague 8, Czech Republic}

\maketitle\abstracts{
We discuss recently introduced scale-free Einstein equations, where
the information from their trace part is lost. These equations are
classically equivalent to General Relativity, yet the Newton constant
becomes a constant of integration or a global dynamical degree of
freedom. Thus, from the point of view of standard quantization, this
effective Newton constant is susceptible to quantum fluctuations.
This is similar to what happens to the cosmological constant in the
unimodular gravity where the trace part of the Einstein equations
is lost in a different way. Using analogy with the Henneaux-Teitelboim
covariant action for the unimodular gravity, we consider different
general-covariant actions resulting in these dynamics. This setup
allows one to formulate the Heisenberg uncertainty relations for the
Newton constant and canonically conjugated quantities. Unexpectedly,
one of such theories also promotes the Planck's quantum constant to
a global degree of freedom, which is subject to quantum fluctuations.
Following analogy with the unimodular gravity, we discuss non-covariant
``unimatter'' and ``unicurvature'' gravities describing the scale-free
Einstein equations. Finally, we show that in some limit of the Yang-Mills
gauge theory a ``frozen'' axion-like field can emulate the gravitational
Newton constant or even of the quantum Planck constant.}
\section{Introduction and Main Idea }
For the last six billion years our Universe has been expanding with an acceleration.
This phenomenon is well described by the gravitational action\footnote{We use $\left(+,-,-,-\right)$ signature convention and units where
$c=1$, while in most parts of the text $\hbar=1$ and we restore
$\hbar$ only to stress the quantum nature of a formula.} 
\begin{equation}
S\left[g\right]=-\frac{1}{16\pi G_{N}}\int d^{4}x\sqrt{-g}\left(R+2\Lambda\right)\,,\label{eq:EH_action}
\end{equation}
where the cosmological constant $\Lambda$ has inverse dimensions
to the Newton constant $G_{N}$, while their product is inexplicably
tiny: 
\begin{equation}
G_{N}\Lambda\sim10^{-122}\,.\label{eq:122}
\end{equation}
The origin of this extremely small number remains a mystery. In particular,
it is not clear how to keep this product small when taking into account
quantum corrections to $\Lambda$ and $G_{N}$. But what if these
constants are in fact dynamical quantities? In that case either some
dynamics may enforce (\ref{eq:122}) or quantum cosmology (possibly
together with anthropic reasoning) may help to select only those initial
values which are close to (\ref{eq:122}). The minimal number
of corresponding degrees of freedom is achieved when $\Lambda$ and/or $G_{N}$ 
are global degrees of freedom. Such degrees of freedom are space-independent
-- they remain constant along every Cauchy surface for arbitrary
3+1 foliation of spacetime. The simplest example of a global degree
of freedom is provided by the so-called unimodular gravity (UG). Indeed,
since Einstein's paper \cite{Einstein:1919gv} it is well known that
trace-free equations
\begin{equation}
G_{\mu\nu}-\frac{1}{4}g_{\mu\nu}G=8\pi G_{N}\left(T_{\mu\nu}-\frac{1}{4}g_{\mu\nu}T\right)\,,\label{eq:Unimod}
\end{equation}
are equivalent to General Relativity (GR) equipped with $\Lambda$
being a constant of integration, or \emph{a global} degree of freedom,
see e.g.\cite{Anderson:1971pn,Weinberg:1988cp,Unruh:1988in,Unruh:1989db,Ellis:2010uc,Henneaux:1989zc}.
This can be checked by applying covariant derivative $\nabla^{\mu}$
to both sides of this equation and consequently using the Bianchi
identity along with the assumed conservation of the total energy-momentum
tensor (EMT). These trace-free Einstein equations (\ref{eq:Unimod})
are invariant under vacuum shifts of the total EMT
\begin{equation}
T_{\mu\nu}\rightarrow T_{\mu\nu}+c\,g_{\mu\nu}\,,\label{eq:vacuum_shifts}
\end{equation}
with $c=c\left(x^{\mu}\right)$, and in particular with $c=const$. 

In \cite{Jirousek:2020vhy} we promoted the Newton constant to a
similar global degree of freedom. This is achieved again by loosing
the information contained in the trace of the Einstein equations.
Namely, instead of the trace-free equations (\ref{eq:Unimod}) one can write
\emph{normalized} or \emph{trace-trivial} equations 
\begin{equation}
\frac{G_{\mu\nu}}{G}=\frac{T_{\mu\nu}}{T}\,,\label{eq:normalized-2}
\end{equation}
or in a more regular form 
\begin{equation}
G_{\alpha\beta}T_{\mu\nu}=G_{\mu\nu}T_{\alpha\beta}\,.\label{eq:more_regular}
\end{equation}
These equations are \emph{scale-free }and invariant with respect to
rescaling
\begin{equation}
T_{\mu\nu}\rightarrow c\,T_{\mu\nu}\,,\label{eq:rescale}
\end{equation}
by a constant, or even an arbitrary function $c\left(x^{\mu}\right)$.
In \cite{Jirousek:2020vhy} we demonstrated that equations (\ref{eq:normalized-2})
(or (\ref{eq:more_regular})) are equivalent to the usual GR, however,
now $G_{N}$ becomes a constant of integration. Similarly to the unimodular
case, this equivalence follows from the Bianchi identity and assumed
EMT conservation. 
\section{Unimodular Gravity}

Let us first briefly recall UG. A simple covariant action\footnote{For a higher-derivative and Weyl-invariant formulation of UG see \cite{Jirousek:2018ago}. }
for UG was introduced by Henneaux and Teitelboim in \cite{Henneaux:1989zc}
\begin{equation}
S\left[g,W,\Lambda\right]=-\frac{1}{16\pi G_{N}}\int d^{4}x\sqrt{-g}\left[R+2\Lambda\left(1-\nabla_{\mu}W^{\mu}\right)\right]\,.\label{eq:vector_for_CC}
\end{equation}
For every foliation of spacetime, variation with respect to the spatial
components $W^{i}$ yields a constraint $\partial_{i}\Lambda=0$.
Hence, $\Lambda$ is a global quantity--is space independent. Following
``quantization without tears'' by Faddeev-Jackiw \cite{Faddeev:1988qp},
one can use this constraint in the action to infer that the energy
density $\varepsilon_{\Lambda}=\Lambda/\left(8\pi G_{N}\right)$ is
a canonical momentum conjugated to another global quantity often called
cosmic time
\begin{equation}
\tau\left(t\right)=\int_{\Sigma}d^{3}\mathbf{x}\sqrt{-g}\,W^{t}\left(t,\mathbf{x}\right)\,.\label{eq:cosmic_time}
\end{equation}
Variation of (\ref{eq:vector_for_CC}) with respect to $\Lambda$
gives $\nabla_{\mu}W^{\mu}=1$. For $W^{i}$ vanishing at the boundary
(e.g. at the spatial infinity) the last equation implies that $\tau\left(t\right)$
measures four-volume $\Omega$ of space-time between Cauchy hypersurfaces
$\Sigma_{2}$ and $\Sigma_{1}$: 
\begin{equation}
\tau\left(t_{2}\right)-\tau\left(t_{1}\right)=\int_{\Omega}d^{4}x\sqrt{-g}\,.\label{eq:4volume}
\end{equation}
As $\varepsilon_{\Lambda}$ and $\tau$ are canonically conjugated,
one can use the correspondence principle of quantization and find
that they are subject to the Heisenberg uncertainty relation 
\begin{equation}
\delta\varepsilon_{\Lambda}\times\delta\tau\geq\frac{\hbar}{2}\,,\label{eq:Energy_time_uncertainty}
\end{equation}
similar to the energy-time uncertainty relation. This uncertainty
relation can be also written as
\begin{equation}
\delta\Lambda\times\delta\int_{\Omega}d^{4}x\sqrt{-g}\geq4\pi\,\ell_{Pl}^{2}\,,\label{eq:uncertainty_volume}
\end{equation}
where $\ell_{Pl}=\sqrt{\hbar G_{N}}$ is the Planck length. These
unavoidable quantum fluctuations are the main difference of UG from
the usual GR. This property is insensitive to higher order curvature
invariants appearing in effective action for gravity. These fluctuations
may be relevant close to singularities. Except of this phenomenon,
perturbative ``unimodular'' gravity is equivalent to GR also in
quantum realm, for recent discussions see e.g. \cite{Alvarez:2005iy,Fiol:2008vk,Smolin:2009ti,Eichhorn:2013xr,Saltas:2014cta,Padilla:2014yea,Bufalo:2015wda,Alvarez:2015sba,Alvarez:2015pla,Percacci:2017fsy,Ardon:2017atk,Herrero-Valea:2018ilg,Herrero-Valea:2020xaq,deBrito:2020rwu}.
There are only two ways to claim that (\ref{eq:uncertainty_volume})
is not applicable. Namely, either $\int_{\Omega}d^{4}x\sqrt{-g}$
is not well defined, say it is divergent, or one is ready to violate
the main postulate of the canonical quantization that the Poisson
bracket for the canonically conjugated quantities are directly mapped
into commutators of the corresponding operators. Another conceptual
issue with a potential to undermine the usefulness of (\ref{eq:uncertainty_volume})
is the following. It may be difficult for a local observer to measure
a total space-time volume between Cauchy hypersurfaces, as this involves
events with space-like separation. In this case, four volume could
have arbitrary large fluctuations, provided they are due to causally
disconnected regions. These arbitrary large fluctuations of four volume
would imply that the observer can be at an eigenstate of $\Lambda$. 

To illustrate the importance of the uncertainty relation let us take
closed radiation-dominated Friedmann Universe
\begin{equation}
ds^{2}=a^{2}\left(\eta\right)\left[d\eta^{2}-d\chi^{2}+\sin^{2}\chi\left(d\theta^{2}+\sin^{2}\theta\,d\phi^{2}\right)\right]\,,\label{eq:Radiation_dominated_U}
\end{equation}
where $a\left(\eta\right)=a_{m}\sin\eta,$ so that conformal time
$\eta$ and radial coordinate $\chi$ both belong to the interval $\left(0,\pi\right)$.
The total four volume for this universe is 
\begin{equation}
\Omega_{tot}=\int_{0}^{\pi}d\eta\int d^{3}x\sqrt{-g}=4\pi a_{m}^{4}\int_{0}^{\pi}\sin^{4}\eta d\eta\,\int_{0}^{\pi}\sin^{2}\chi d\chi=\frac{3\pi^{3}}{4}a_{m}^{4}\,.\label{eq:4_volume_radiation_dom}
\end{equation}
Quasiclassical approximation requires that at least $\delta\tau<\Omega_{tot}$.
Hence, under the assumption of quasiclassical evolution, equation (\ref{eq:Energy_time_uncertainty})
yields the bound 
\begin{equation}
\varepsilon_{\Lambda}>\delta\varepsilon_{\Lambda}>\frac{4}{3\pi^{3}}\,\frac{\hbar}{a_{m}^{4}}\,.\label{eq:bound_radiation_universe}
\end{equation}
For a large Universe this is a tiny bound independent of the composition
and structure of the radiation. It is useful to compare (\ref{eq:bound_radiation_universe})
with the minimal energy density $\varepsilon_{m}=3/\left(8\pi G_{N}a_{m}^{2}\right)$.
Hence, $\varepsilon_{m}\gg\delta\varepsilon_{\Lambda}$ provided $a_{m}\gg\ell_{Pl}$. 
\section{Action for Scale-Free Gravity, Changing Gravity \label{sec:Action-for-Einstein-transverse} }

Following the Henneaux-Teitelboim formulation of the UG (\ref{eq:vector_for_CC})
discussed above, it is easy to write a similar action to promote the
Newton constant to a global degree of freedom\footnote{cf. \cite{Kaloper:2015jra,Kaloper:2018kma} written for local version
of the vacuum energy sequester \cite{Kaloper:2013zca}}: 
\begin{equation}
S\left[g,C,\alpha\right]=\frac{1}{2}\int d^{4}x\sqrt{-g}\left(\nabla_{\mu}C^{\mu}-R\right)\alpha\,.\label{eq:Vector_for_G}
\end{equation}
Variation with respect to $C^{\mu}$ yields $\partial_{\mu}\alpha=0$
while variation with respect to the metric gives 
\begin{equation}
\alpha\,G_{\mu\nu}=T_{\mu\nu}\,.\label{eq:alpha_Einstein-1}
\end{equation}
Thus $G_{N}=\left(8\pi\alpha\right)^{-1}$ becomes a constant of integration.
In complete analogy with Henneaux-Teitelboim formulation of UG discussed
in the previous section, we obtain that $\alpha$ is a canonical momentum
conjugated to the global quantity 
\begin{equation}
\varrho\left(t\right)=\frac{1}{2}\int_{\Sigma}d^{3}\mathbf{x}\sqrt{-g}\,C^{t}\left(t,\mathbf{x}\right)\,.\label{eq:global_curvature_charge}
\end{equation}
For appropriate boundary conditions on $C^{i}$, due to the constraint,
$\nabla_{\mu}C^{\mu}=R$, this global quantity measures the integrated
Ricci scalar 
\begin{equation}
\varrho\left(t_{2}\right)-\varrho\left(t_{1}\right)=\frac{1}{2}\int_{\Omega}d^{4}x\sqrt{-g}\,R\,.\label{eq:4volume_R}
\end{equation}
 For canonically conjugated pair $\left(\varrho,\alpha\right)$
the Heisenberg uncertainty relation reads $\delta\varrho\times\delta\alpha\geq\hbar/2$.
One can write it using observable
quantities and the Planck length $\ell_{Pl}=\sqrt{\hbar G_{N}}$ as
\begin{equation}
\frac{\delta G_{N}}{G_{N}}\times\frac{\delta\int_{\Omega}d^{4}x\sqrt{-g}\,R}{\ell_{Pl}^{2}}\geq8\pi\,.\label{eq:uncertainty_G_alpha}
\end{equation}
Again this inequality should have rather nontrivial consequences close
to singularities. Quasiclassical description implies that 
\begin{equation}
8\pi G_{N}\left|\int_{\Omega}d^{4}x\sqrt{-g}T\right|\gg\delta\int_{\Omega}d^{4}x\sqrt{-g}\,R\,.\label{eq:quasiclassical_T}
\end{equation}
Thus, there is a lower bound on fluctuations of the Newton constant
\begin{equation}
\frac{\delta G_{N}}{G_{N}}\gg\hbar\left|\int_{\Omega}d^{4}x\sqrt{-g}\,T\right|^{-1}\,.\label{eq:Bound_on_G}
\end{equation}
Hence, conformal anomaly plays a crucial role for the magnitude of
fluctuations of $\delta G_{N}$. On the other hand, most of the fields
in standard model are conformal for hight temperatures. But on top
of conformal anomaly there can be vacuum energy $\varepsilon_{\Lambda}$
as the source of $T$. In that case (\ref{eq:Bound_on_G}) implies
that 
\begin{equation}
\frac{\delta G_{N}}{G_{N}}\gg\frac{\hbar}{4\varepsilon_{\Lambda}}\left(\int_{\Omega}d^{4}x\sqrt{-g}\right)^{-1}\,.\label{eq:Bound_on_G/Lambda}
\end{equation}
The rather weak lower bounds (\ref{eq:Bound_on_G}) and (\ref{eq:Bound_on_G/Lambda})
is a novel extension of material presented in \cite{Jirousek:2020vhy}. 
\section{Action for Scale-Free Gravity, Changing Matter\label{sec:Action-for-energy-transverse}}
Interestingly there is another way to write an action for the Newton
constant as a global degree of freedom. Namely, one can preserve the
Einstein-Hilbert action, but change the usual action for matter fields
$\Phi_{m}$, 
\begin{equation}
S_{0}\left[g,\Phi_{m}\right]=\int d^{4}x\sqrt{-g}\,\mathcal{L}_{m}\,,\label{eq:usual_Action_matter}
\end{equation}
to one (c.f. \cite{Carroll:2017gqo}) similar to (\ref{eq:Vector_for_G}):
\begin{equation}
S\left[g,\beta,L,\Phi_{m}\right]=\int d^{4}x\sqrt{-g}\,\beta\left(\mathcal{L}_{m}-\nabla_{\mu}L^{\mu}\right)\,.\label{eq:rescaled_matter_action}
\end{equation}
In this formulation of the theory, on the right hand side of the Einstein
equations one obtains a rescaled EMT for matter fields $\Phi_{m}$
\begin{equation}
T_{\mu\nu}=\frac{2}{\sqrt{-g}}\frac{\delta S}{\delta g^{\mu\nu}}=\beta\,T_{\mu\nu}^{\left(m\right)},\label{eq:rescaled_EMT}
\end{equation}
where 
\begin{equation}
T_{\mu\nu}^{\left(m\right)}=\frac{2}{\sqrt{-g}}\frac{\delta S_{0}}{\delta g^{\mu\nu}}=\frac{2}{\sqrt{-g}}\frac{\delta\left(\sqrt{-g}\mathcal{L}_{m}\right)}{\delta g^{\mu\nu}}\,.\label{eq:usual_EMT_matter}
\end{equation}
Variation with respect to $L^{\mu}$ yields $\partial_{\mu}\beta=0$.
Hence, the effective Newton constant, $\bar{G}_{N}$, is just a rescaling
of the constant from the Einstein-Hilbert action
\begin{equation}
\bar{G}_{N}=G_{N}\beta\,.\label{eq:rescaled_G_N}
\end{equation}
Following the same path as in (\ref{eq:cosmic_time}) and (\ref{eq:global_curvature_charge}),
the global dynamical degree of freedom canonically conjugated to $\beta$
is 
\begin{equation}
I\left(t\right)=-\int_{\Sigma}d^{3}\mathbf{x}\sqrt{-g}\,L^{t}\left(t,\mathbf{x}\right)\,.\label{eq:Matter_action_charge}
\end{equation}
Under appropriate boundary conditions, due to $\nabla_{\mu}L^{\mu}=\mathcal{L}_{m}$,
this variable measures the matter action between the Cauchy hypersurfaces
$\Sigma_{2}$ and $\Sigma_{1}$
\begin{equation}
I\left(t_{2}\right)-I\left(t_{1}\right)=-\int_{\Omega}d^{4}x\sqrt{-g}\,\mathcal{L}_{m}\,.\label{eq:Matter_Action}
\end{equation}
Applying again the Heisenberg uncertainty relation to the canonical
pair $\left(I,\beta\right)$ one obtains $\delta I\times\delta\beta\geq\hbar/2$,
which can be written in terms of observables as 
\begin{equation}
\delta\bar{G}_{N}\times\delta\int_{\Omega}d^{4}x\sqrt{-g}\,\mathcal{L}_{m}\geq\frac{1}{2}\ell_{Pl}^{2}\,,\label{eq:matter_uncertainty_Realtion}
\end{equation}
where the Planck length $\ell_{Pl}=\sqrt{\hbar G_{N}}$ corresponds
to $G_{N}$ from the Einstein-Hilbert action. 

More interestingly, the introduction of $\beta$ also rescales the
commutation relations for usual matter. For instance, for the usual
scalar field $\phi$ the canonical momentum gets rescaled similarly
to the EMT: 
\begin{equation}
\pi=\beta\pi^{\left(m\right)}=\beta\sqrt{-g}\frac{\partial\mathcal{L}_{m}}{\partial\dot{\phi}}\,.\label{eq:canonical_momentum_rescaled}
\end{equation}
Thus, the canonical commutator relation, $\left[\phi\left(\mathbf{x}\right),\pi\left(\mathbf{y}\right)\right]=i\hbar\delta\left(\mathbf{x}-\mathbf{y}\right)$,
implies that usual commutator gets rescaled as well
\begin{equation}
\left[\phi\left(\mathbf{x}\right),\pi^{\left(m\right)}\left(\mathbf{y}\right)\right]=\frac{i\hbar}{\beta}\delta\left(\mathbf{x}-\mathbf{y}\right)\,.\label{eq:rescaled_canonical_commutator}
\end{equation}
Hence, the effective Planck constant is
\begin{equation}
\bar{\hbar}=\frac{\hbar}{\beta}\,.\label{eq:rescaled_Planck_constant}
\end{equation}
Interestingly, the Planck length (and time) $\ell_{Pl}=\sqrt{\hbar G_{N}}$
remains invariant under this rescaling, as $\bar{\hbar}\bar{G}_{N}=\hbar G_{N}$.
Using this invariance one can write (\ref{eq:matter_uncertainty_Realtion})
as 
\begin{equation}
\frac{\delta\bar{G}_{N}}{\bar{G}_{N}}\times\delta\int_{\Omega}d^{4}x\sqrt{-g}\,\mathcal{L}_{m}\geq\frac{1}{2}\bar{\hbar}\,.\label{eq:relative_Newton_fluct}
\end{equation}
Following the same logic as in (\ref{eq:Bound_on_G}) one obtains
a new lower bound 
\begin{equation}
\frac{\delta\bar{G}_{N}}{\bar{G}_{N}}\geq\frac{1}{2}\bar{\hbar}\left|\int_{\Omega}d^{4}x\sqrt{-g}\,\mathcal{L}_{m}\right|^{-1}\,.\label{eq:Bound_on_G_eff}
\end{equation}
Even more unorthodox, one can interpret the uncertainty relation $\delta I\times\delta\beta\geq\hbar/2$
as uncertainty in the effective Planck constant. Indeed, through (\ref{eq:rescaled_Planck_constant})
fluctuations $\delta\beta$ correspond to $\delta\bar{\hbar}=\hbar\delta\beta/\beta^{2}$
so that 
\begin{equation}
\delta\bar{\hbar}\times\delta\int_{\Omega}d^{4}x\sqrt{-g}\,\mathcal{L}_{m}\geq\frac{1}{2}\bar{\hbar}^{2}\,,\label{eq:uncertainty_for_Planck}
\end{equation}
with the corresponding lower bound on fluctuations 
\begin{equation}
\frac{\delta\bar{\hbar}}{\bar{\hbar}}\geq\frac{1}{2}\bar{\hbar}\left|\int_{\Omega}d^{4}x\sqrt{-g}\,\mathcal{L}_{m}\right|^{-1}\,.\label{eq:Lower_Bound_Fluct_hbarbar}
\end{equation}
For many theories the Lagrangian density is vanishing on equations
of motion. However, similarly to (\ref{eq:Bound_on_G/Lambda}) one
can include vacuum energy $\varepsilon_{\Lambda}$ into $\mathcal{L}_{m}$. 
\section{Unimodular, Unicuravature and Unimatter }

Fixing a gauge and a class of coordinates in the UG action (\ref{eq:vector_for_CC})
 such that 

\begin{equation}
W^{\mu}=\delta_{t}^{\mu}\frac{t}{\sqrt{-g}}\,,\label{eq:fixed_W}
\end{equation}
\emph{before} performing the variation ensues
\begin{equation}
\int d^{4}x\,\Lambda\left[1-\sqrt{-g}\,\right]\,.\label{eq:fixed_W_constraint}
\end{equation}
This is a usual non-covariant formulation of the UG \cite{Buchmuller:1988wx,Buchmuller:1988yn,vanderBij:1981ym,Alvarez:2006uu}.
Interestingly, this formulation still results in the seemingly covariant
traceless Einstein equations (\ref{eq:Unimod}). In \cite{Jirousek:2020vhy}
we demonstrated that one can do the same to obtain non-covariant formulations
for theories with the globally dynamical Newton (\ref{eq:Vector_for_G})
and Planck constants (\ref{eq:rescaled_matter_action}). 

Indeed, similarly to (\ref{eq:fixed_W}) one can fix
\begin{equation}
C^{\mu}=\pm\delta_{t}^{\mu}\frac{t}{\sqrt{-g}}\,M_{\alpha}^{2}\,,\label{eq:fixed_C}
\end{equation}
in the action, \emph{before} variation. Here, $"+"$ corresponds to
the positive Ricci scalar, while $"-"$ to the negative one, while
$M_{\alpha}$ is some mass scale introduced for dimensional reasons.
Then, in units where $M_{\alpha}=1$, the action takes a non-covariant
form
\begin{equation}
S\left[g,\alpha\right]=\frac{1}{2}\int d^{4}x\left(\pm1-\sqrt{-g}R\right)\alpha\,.\label{eq:unicurvature_action}
\end{equation}
Thus instead of the ``unimodular'' constraint $\sqrt{-g}=1$ we
have a ``unicurvature'' condition $\sqrt{-g}R=\pm1$. In \cite{Jirousek:2020vhy}
we provided a proof that the action (\ref{eq:unicurvature_action})
together with the usual matter action (\ref{eq:usual_Action_matter})
describe same trace-trivial equations (\ref{eq:normalized-2}).
Thus, along with the well-known unimodular gravity one can study the novel ``unicurvature''
gravity (\ref{eq:unicurvature_action}). Here it was crucial that
the gauge fixing does not affect the matter sector so that the total
EMT is conserved. 

Finally, one can fix 
\begin{equation}
L^{\mu}=\pm\delta_{t}^{\mu}\frac{t}{\sqrt{-g}}\,M_{\beta}^{4}\,,\label{eq:fixed_L}
\end{equation}
\emph{before} varying the action (\ref{eq:rescaled_matter_action}).
Here $M_{\beta}$ is again some mass scale in units of which the action
takes an unusual non-covariant form 
\begin{equation}
S\left[g,\beta,\Phi_{m}\right]=\int d^{4}x\,\beta\left(\sqrt{-g}\mathcal{L}_{m}\mp1\right)\,.\label{eq:unimatter}
\end{equation}
Variation with respect to $\beta$ results in $\sqrt{-g}\mathcal{L}_{m}=\pm1$
which by analogy with the unimodular constraint $\sqrt{-g}=1$ can
be called ``unimatter'' condition. In \cite{Jirousek:2020vhy}
we provided a proof that action (\ref{eq:unimatter}) accompanied
with the usual Einstein-Hilbert action again describe the  same trace-trivial
equations (\ref{eq:normalized-2}). Thus, this construction can be
called ``unimatter'' gravity. The key point of the proof in\cite{Jirousek:2020vhy}
is that the transformation properties of (\ref{eq:unimatter}) with
respect to diffeomorphisms imply that $\nabla^{\nu}\left(\beta T_{\mu\nu}^{\left(m\right)}\right)=-\mathcal{L}_{m}\partial_{\mu}\beta$.
Then the constancy of $\beta$ ensues from the Bianchi identity. 
\section{Frozen Axions}

In \cite{Hammer:2020dqp} it was showed\footnote{cf. \cite{Kimpton:2012rv}},
that vector field $W^{\mu}$ in UG can be exchanged in (\ref{eq:vector_for_CC})
with a more convenient Chern-Simons current of a (non-abelian or abelian)
gauge field $A_{\mu}$. In this way instead of $\nabla_{\mu}W^{\mu}$
one can plug in $F_{\alpha\beta}F^{\star\alpha\beta}$, where the
gauge field strength is $F_{\mu\nu}=D_{\mu}A_{\nu}-D_{\nu}A_{\mu}$
with the covariant derivative $D_{\mu}=\nabla_{\mu}+iA_{\mu}$, while
the Hodge dual is defined as usual 
\begin{equation}
F^{\star\alpha\beta}=\frac{1}{2}\,\frac{\epsilon^{\alpha\beta\mu\nu}}{\sqrt{-g}}\, F_{\mu\nu}\equiv\frac{1}{2}\, E^{\alpha\beta\mu\nu}\, F_{\mu\nu}\,.\label{eq:dual}
\end{equation}
In the absence of the usual kinetic term $-F_{\alpha\beta}F^{\alpha\beta}/4{g}^{2}$,
variation of the action with respect to $A_{\mu}$ forces $\Lambda$
to be constant. It is easy to further extend \cite{Hammer:2020dqp}
this action to resemble the one of the usual axion, cf. \cite{Wilczek:1983as}
\begin{equation}
S\left[g,A,\theta\right]=\int d^{4}x\sqrt{-g}\left[-\frac{R}{16\pi G_{N}}+\frac{1}{2}\left(\partial\theta\right)^{2}+\frac{\theta}{f_{\Lambda}}F_{\alpha\beta}F^{\star\alpha\beta}-V_{\lambda}\left(\theta\right)\right]\,,\label{eq:axion_for_Lambda}
\end{equation}
where now $\theta$ is a canonically normalized pseudoscalar, while
$f_{\Lambda}$ is some mass scale emulating axion decay constant and
$V_{\lambda}\left(\theta\right)$ is a ``potential''. We would like
to stress again, that it is the absence of the usual kinetic term
for the gauge field $-F_{\alpha\beta}F^{\alpha\beta}/4{g}^{2}$
which freezes $\theta$ to be constant. This gives hopes to embed
unimodular gravity in some more usual high energy system. Indeed,
this action can appear from the usual axion and Yang-Mills construction
in the dynamical regime when one can neglect the usual kinetic term
$-F_{\alpha\beta}F^{\alpha\beta}/4{g}^{2}$ for the gauge field.
Naively, this setup corresponds to a confinement or an infinitely
strong coupling ${g}\rightarrow\infty$ in IR. 

In \cite{Jirousek:2020vhy} we proposed that the same procedure can
be done for theories with the globally dynamical Newton (\ref{eq:Vector_for_G})
and Planck constants (\ref{eq:rescaled_matter_action}). Namely (\ref{eq:Vector_for_G}) can be extended to 
\begin{equation}
S\left[g,\mathcal{A},\nu\right]=\int d^{4}x\sqrt{-g}\left[-\frac{1}{2}\nu^{2}R+\frac{1}{2}\left(\partial\nu\right)^{2}+\frac{\nu}{f_{\alpha}}\mathcal{F}_{\gamma\sigma}\mathcal{F}^{\star\gamma\sigma}-V_{\alpha}\left(\nu\right)\right]\,,\label{eq:axion_for_Newton}
\end{equation}
where now $\nu$ is another frozen ``axion'', which is constant
on all solutions, while $f_{\alpha}$ is some mass scale and $V_{\alpha}\left(\nu\right)$
is a corresponding potential. The presence of non-minimal coupling
to gravity breaks shift-symmetry. Thus, potential and the ensuing
cosmological constant appear naturally in this setup. This action is
written in the Jordan frame and we remind the reader that the standard
Einstein-Hilbert term is absent. Connecting to (\ref{eq:Vector_for_G})
one notices that $\alpha=\nu^{2}$, so that the effective Newton constant
is given by $G_{N}=1/(8\pi\nu^{2})$. 

Completely analogously to the previous cases, one can extend (\ref{eq:rescaled_matter_action})
to
\begin{equation}
S\left[g,\eta,\mathcal{A},\Phi_{m}\right]=\int d^{4}x\sqrt{-g}\left[\frac{1}{2}\left(\partial\eta\right)^{2}-V_{\beta}\left(\eta\right)+\frac{\eta^{2}}{M_{m}^{2}}\mathcal{L}_{m}-\frac{\eta}{f_{\beta}}\mathcal{F}_{\mu\nu}\mathcal{F}^{\star\mu\nu}\right]\,,\label{eq:axion_unimatter}
\end{equation}
where $\eta$ is again a frozen ``axion'', which is constant on
all solutions. This frozen ``axion'' \emph{universally} couples
to the Lagrangian $\mathcal{L}_{m}$ of all other matter fields $\Phi_{m}$.
This coupling is quadratic, as this is the lowest possible order which
does not introduce ghosts. Due to this coupling the shift-symmetry
is broken. Hence, potential $V_{\beta}\left(\eta\right)$ with the
corresponding cosmological constant are naturally incorporated in
this construction. The usual dimensions of $\eta$ are restored by
introducing two mass scales $M_{m}$ and $f_{\beta}$. Here gravity 
is described by the standard Einstein-Hilbert action. Connecting to (\ref{eq:rescaled_matter_action})
one notices that $\beta=\eta^{2}/M_{m}^{2}$, so that the effective
Newton constant (\ref{eq:rescaled_G_N}) is given by $\bar{G}_{N}=G_{N}\eta^{2}/M_{m}^{2}$
while the effective Planck constant (\ref{eq:rescaled_Planck_constant})
is $\bar{\hbar}=\hbar M_{m}^{2}/\eta^{2}$. 

Finally, it is worth mentioning that combining (\ref{eq:axion_unimatter})
or (\ref{eq:axion_for_Newton}) with the (\ref{eq:axion_for_Lambda})
provides an ``axionic'' setup for the vacuum energy sequestering \cite{Kaloper:2013zca,Kaloper:2015jra} and brings
it closer to the usual particle physics models. 
\section{Conclusions }

In \cite{Jirousek:2020vhy} we proposed theories where the effective
Newton constant and Planck constants are global dynamical degrees
of freedom. In this proceeding we first presented scale-free or scale
trivial Einstein equations where the Newton constant is the constant
of integration. In Section II we recall covariant formulation of the
unimodular gravity (UG) due to Henneaux and Teitelboim \cite{Henneaux:1989zc}.
In particular we stressed the presence of the unavoidable uncertainty
relation which distinguishes quasiclassical UG from GR. Then, in Section
III and IV we presented two distinct extensions of Henneaux and Teitelboim
construction to describe global dynamics of $G_{N}$ and $\hbar$,
(\ref{eq:Vector_for_G}) and (\ref{eq:rescaled_matter_action}) respectively.
Extending \cite{Jirousek:2020vhy}, we derived new lower bounds
on quantum fluctuations of effective $G_{N}$, see (\ref{eq:Bound_on_G}), (\ref{eq:Bound_on_G_eff}) and (\ref{eq:Lower_Bound_Fluct_hbarbar})
for effective $\hbar$. Then, in Section V we discussed non-covariant
formulations of such theories: so-called ``unimatter'' (\ref{eq:unimatter})
and ``unicurvature'' (\ref{eq:unicurvature_action}) gravities.
Finally, in Section VI we considered how one can embed covariant formulations
of such theories into axion and a confined Yang-Mills gauge theory. 

There remain a lot of issues to understand better in such theories.
In particular one can mention: minisuperspace quantum cosmology; canonical
analysis and quantization for axionic construction; boundary terms
and matching conditions; and finally global degrees of freedom in
the presence of horizons and more importantly close to the end-of-time
singularities of GR. It would be exciting, if quantum mechanics of
such global degrees of freedom could solve the global end/beginning
of time problems in GR. 
\section*{Acknowledgments}
It is a great pleasure to thank Pavel Jirou\v{s}ek, Keigo Shimada and
Masahide Yamaguchi for a fruitful collaboration resulting in \cite{Jirousek:2020vhy}.
Special thanks are for Pavel Jirou\v{s}ek who carefully red the first version of the text. The work of A.\,V. on \cite{Jirousek:2020vhy} and on this proceedings
contribution is supported by the funds from the European Regional
Development Fund and the Czech Ministry of Education, Youth and Sports
(M\foreignlanguage{american}{\v{S}}MT): Project CoGraDS - CZ.02.1.01/0.0/0.0/15\_003/0000437. 

\end{document}